\documentclass[onecolumn, 12pt, epsf]{article}
\usepackage{amsmath, amsthm, amssymb, graphicx,subfig,indentfirst,deluxetable,feynmf}
\begin{document}
\def\frc#1#2{{\textstyle{#1 \over #2}}}
\begin{center}
\LARGE Stationary Point of the Hilbert\\
Action Principle \vspace{1.5in}\\
\large James W. York \vspace{0.25in}\\
\normalsize Department of Physics\\
North Carolina State University\\
Raleigh, NC, 27695-8202 \vspace{0.25in}\\
November $\rm{11}^{\rm{th}}$, 2015 \vspace{0.7in}\\
\begin{abstract} The Hilbert action principle for the Einstein equations has two boundary terms that must be subtracted in prder to obtain a true stationary point.  In this paper, I obtain both of them.
\end{abstract}
\end{center}
Assume that spacetime $V$ is topologically $\Sigma \times \mathbb{R}$ where $\Sigma$ is a spacelike Cauchy slice.  The splittings of the spacetime curvature tensor $R_{\mu\nu\alpha\beta}$ works as follows\cite{Ricci Calc}.
\begin{align}
R_{ijkl}&=\bar{R}_{ijkl}+\left(K_{ik}K_{jl}-K_{il}K_{jk}\right), &\text{Gauss\label{eq1}}\\
R_{0ijk}&=N\left(\nabla_jK_{ik}-\nabla_kK_{ij}\right),&\text{Codazzi}\\
R_{0i0j}&=N\left(\hat{\partial_0}K_{ij}+K_{ik}K^k_j+\nabla_i\nabla_jNN\right),&\text{Ricci}
\end{align}
where $\hat{\partial}_0=\partial_t-\mathcal{L}_\beta$.  $N$ is the lapse function and $\beta^i$ is the shift vector.  In (\ref{eq1}) $\bar{R}_{ijkl}$ is the Reimann tensor of $\Sigma$ and $\nabla_i$ is the spatial covariant derivative.  In three dimmensions $\bar{R}_{ijkl}$ is equivalent to the Ricci Tensor $\bar{R}_{ij}$. $K_{ij}$ is the extrinsic curvature of $\Sigma$.  Proceeding with the splitting of spacetime curvatures tensors, I find

\begin{equation}
Ric_{\beta\gamma}=R_{\alpha\beta}{}^\alpha{}_\gamma
\end{equation}
\begin{equation}
Ric_{ij}=\bar{R}_{ij}-N^{-1}\hat{\partial}_0K_{ij}+KK_{ij}-2K_{ij}K^k{}_j+N^{-1}\nabla_i\nabla_j N,\\
\end{equation}\\
where $K=g_{ij}K^{ij}$
\begin{equation}
Ric_{00}=N\left(\hat{\partial}_0 K-N K_{ij}K^{ij}\Delta N\right),\\
\end{equation}
with $\Delta =g^{ij}\nabla_i\nabla_j$.\\
\\
The spacetime scalar curvature is $R=g^{\alpha\beta}R_{ic\alpha\beta}$ from which
\begin{equation}
R=N\hat{\partial}_0K-2N^{-1}\Delta N+\left(\bar{R}+K_{ij} K^{ij}- K^2\right)
\end{equation}
and
\begin{eqnarray}
N\sqrt{g}R&=&N\sqrt{g}\left(\bar{R}+K_{ij}K^{ij}-K^2\right)-\partial_t\left(\sqrt{g}K\right)\\
&+&\partial_i\left[\sqrt{g}\left(K\beta^i-g^{ij}\partial_jN\right)\right]. \nonumber
\end{eqnarray}
The tensor $\left(-\partial_t\sqrt{g}k\right)$ was recognized in \cite{york}.  The spatial scalar curvature is
\begin{equation}\label{nine}
\bar{R}=R_{(2)}-\left(\chi_{ij}\chi^{ij}+\chi^2\right)+2\left(\mathcal{L}_s\chi -\alpha^{-1}\bar{g}^{ij}\bar{\nabla}_i\bar{\nabla}_j\alpha\right),
\end{equation}
where $R_{(2)}$ is the scalar curvature of the reference this space $S$ embedded in $\Sigma$.  The metric of $S$ is $\bar{g}_{ij}$ and its covariant derivative is $\bar{\nabla}_i$.  
Assume that $\Sigma$ is foliated by $S$'s, all of which are outside of any sources.  
The analogue of $N$ is $\alpha$, a lapse function associated with the foliation of $\Sigma$ by $S$'s.  
The $S$'s are topologically two-spheres that approach a genuine two-sphere as $r\rightarrow\infty$ in asymptotically Euclidean $\Sigma$'s (AES's).
    Define a unit vector $s^i$ that is the outward-pointing unit normal of the $S$'s.  The extensive curvature of $S$ in $\Sigma$ is $\chi_{ij}$; it satisfies
\begin{equation}
\chi_{ij}=\left(2\alpha\right)^{-1}\left[\partial_r\bar{g}_{ij}+\left(\bar{\nabla}_i s_j +\bar{\nabla}_j s_i\right)\right].
\end{equation}
$\mathcal{L}_N\chi=\mathcal{L}_N\left(\bar{g}^{ij}\chi_{ij}\right)$ has support on AES's because the $S$'s approach a genuine two-sphere in the limit as $r\rightarrow\infty$.
    At asymptotia, the exact form of the Hilbert action principle (with $k=4\pi G$, $G=6.670(4)\times 10^{-8}\rm{gm}^{-1}\rm{cm}^3\rm{sec}^{-2}$ and $c=2.9979250(10)\times10^{10}\rm{cm}\;\rm{sec}^{-1}$) is
\begin{align}\label{eleven}
\left(c^4/4k\right)&\int N\sqrt{g}Rd^4k=(c^4/4k)\int[N\sqrt{g}\left(\bar{R}+K_{ij}K^{ij}-K^2\right)\\
+&\partial_i\left(\sqrt{g}K\beta^i-g^{ij}\partial_jN\right)-\partial_t\left(\sqrt{g}K\right)-\mathcal{L}_N\chi]d^4\chi.\nonumber
\end{align}
The other tensors in (\ref{nine}) have no support as $r\rightarrow\infty$.  Variation of (\ref{eleven}) with respect to $g_{\mu\nu}$ gives as a genuine stationary point
\begin{equation}\label{twelve}
\left(c^4/4k\right)Ein^{\mu\nu}=\left(c^4/4k\right)\left[Ric^{\mu\nu}-\frc12 g^{\mu\nu}R\right]=0.
\end{equation}
$D_\nu Ein^{\mu\nu}\equiv 0$, where $D_\nu$ is the spacetime covariant derivative.
Equation (\ref{twelve}) gives the vacuum Einstein equations.  If sources are present, a term must be added to (\ref{eleven}) to produce stress-energy-momentum tensor $T^{\mu\nu}$ that necessarily satisfies $D_\nu T^{\mu\nu}=$.  This is not a true conservation law except in the presence of one or more Killing vectors $\xi^\mu$ that satisfy
\begin{equation}
\mathcal{L}_\xi g_{\mu\nu}=D_\nu\xi_\mu+D_\mu\xi_\nu=0,
\end{equation}
in which case
\begin{equation}
D_\mu\left(T^{\mu\nu}\xi_\nu\right)=0
\end{equation}
is a true conservation law.

\section*{Acknowledgment}
The author thanks J. David Brown for finding a missing page of his CV.


\begin{thebibliography}{2}

\bibitem{Ricci Calc}
J.A. Schouten, {\em Ricci-Calculus, second edition (1954), Springer-Verlag},\\ Berlin-G{\"o}ttingen-Heidelberg.

\bibitem{york}
J.W. York, Boundary terms in the action principles of general relativity (in honor of the seventy-fifth birthday of John A. Wheeler).  Foundations of Physics 16:249-258 (1986).



\end{thebibliography}
\end{document}